# OPTIMIZED PERFORMANCE EVALUATION OF LTE HARD HANDOVER ALGORITHM WITH AVERAGE RSRP CONSTRAINT

[1]Cheng-Chung Lin, [2]Kumbesan Sandrasegaran, [3]Huda Adibah Mohd Ramli, and [4]Riyaj Basukala

[1,2,3,4] Faculty of Engineering and Information Technology, University of Technology, Sydney, Australia

[[1]Cheng-Chung.Lin, [2]kumbes, [3]HudaAdibah.MohdRamli]@eng.uts.edu.au
[4]Riyaj.Basukala@uts.edu.au

## ABSTRACT

*Hard handover mechanism is adopted to be used in 3GPP Long Term Evolution (3GPP LTE) in order to reduce the complexity of the LTE network architecture. This mechanism comes with degradation in system throughput as well as a higher system delay. This paper proposes a new handover algorithm known as LTE Hard Handover Algorithm with Average Received Signal Reference Power (RSRP) Constraint (LHHAARC) in order to minimize number of handovers and the system delay as well as maximize the system throughput. An optimized system performance of the LHHAARC is evaluated and compared with three well-known handover algorithms via computer simulation. The simulation results show that the LHHAARC outperforms three well-known handover algorithms by having less number of average handovers per UE per second, shorter total system delay whilst maintaining a higher total system throughput.*

## KEYWORDS

*Performance, Handover, Handover algorithm, Optimization, LTE*

## 1. INTRODUCTION

3GPP LTE is a new radio access technology proposed to provide a smooth migration towards Fourth Generation (4G) network [1]. It is designed to increase the capacity, coverage, and speed as compared to the earlier wireless systems [2, 3]. LTE uses Orthogonal Frequency Division Multiple Access (OFDMA), which is a variant of OFDM (Orthogonal Frequency Division Multiplexing), as its access technology in the downlink [4] while single-carrier frequency-division multiple access (SC-FDMA) is the uplink multiple access scheme [5]. OFDMA is a multi-carrier access technology that divides the wide available bandwidth into multiple equally spaced and mutually orthogonal sub-carriers [6]. The smallest transmission unit in the downlink LTE system is known as a Resource Block (RB) that contains 12 sub-carriers (180 kHz total bandwidth) of 1 ms duration [7].

The LTE network architecture consists of three elements: evolved-NodeB (eNodeB), Mobile Management Entity (MME), and Serving Gateway (S-GW) / Packet Data Network Gateway (P-GW). eNodeB performs all radio interface-related functions such as packet scheduling and handover. MME manages mobility, user equipment (UE) identity, and security parameters. S-GW and P-GW are the nodes that terminate the interface towards eUTRAN and Packet Data Network, respectively. There are two interfaces concerned in handovers in eUTRAN which are S1 and X2 interfaces. Both interfaces can be used in handover procedures, but with different purposes. More details about the handover procedures on S1 and X2 interfaces are discussed later.

Handover in LTE is purely hard handover (both S1 and X2 interface handover). The use of hard handover reduces the complexity of the LTE network architecture. However, the hard handover may result inefficient LTE performance (i.e. increasing number of handovers, decreasing system throughput as well as increasing system delay). Therefore, an efficient handover algorithm that can minimize the number of handovers and system delay as well as maximize the system throughput is needed.

A handover algorithm is used for making a handover decision. A handover will be triggered if several conditions specified by a handover algorithm are satisfied. Due to the user's mobility, the conditions of a handover algorithm could vary over time. Therefore it is necessary to determine optimized parameters to ensure efficiency and reliability of a handover algorithm.

A new handover algorithm known as LTE Hard Handover Algorithm with Average RSRP Constraint (LHHAARC) that can efficiently reduce the number of handovers, minimizing the total system delay and maximizing the total system throughput is proposed in this paper. The LHHAARC algorithm is evaluated and compared with three well known handover algorithms using optimized handover parameters under three different speed (3, 30, 120 km/hr) scenarios.

The rest of this paper is organized as follows: Section 2 reviews on the related handover studies followed by detailed descriptions of the well-known and proposed handover algorithms in Section 3. The metrics used for performance evaluation and simulation environment are discussed in Section 4 and Section 5 respectively. Section 6 contains results of the optimization and performance evaluation and conclusions are summarized in Section 7.

## 2. HANDOVER TECHNIQUES

Handover refers to the transfer of a user's connection from one radio channel to another (can be the same or different cell) [8]. Handover can be categorized as hard handover [9] and soft handover [10] also known as Break-Before-Connect (BBC) and Connect (Entry)-Before-Break (CBB), respectively. Soft and hard handover followed by handover in LTE are discussed in the following subsections.

### 2.1. Soft Handover – Connect-Before-Break Handover

Soft handover is a category of handover procedures where the radio links are added and abandoned in such manner that the UE always keeps at least one radio link to the UTRAN [8]. Soft and softer handover were introduced in WCDMA architecture. There is a centralized controller called Radio Network Controller (RNC) to perform handover control for each UE in the architecture of WCDMA. It is possible for a UE to simultaneously connect to two or more cells (or cell sectors) during a call [11]. If the cells the UE connected are from the same physical site, it is referred as softer handover. In handover aspect, soft handover is suitable for maintaining an active session, preventing voice call dropping, and resetting a packet session. However, the soft handover requires much more complicated signalling, procedures and system architecture such as in the WCDMA network.

### 2.2. Hard Handover – Break-Before-Connect Handover

Hard handover is a category of handover procedures where all the old radio links in the UE are abandoned before the new radio links are established [8]. The hard handover is commonly used when dealing with handovers in the legacy wireless systems. The hard handover requires a user to break the existing connection with the current cell (source cell) and make a new connection to the target cell.

### 2.3. Handover in LTE

There are two types of handover procedure in downlink LTE for UEs in active mode (Active mode means the UE is transmitting/receiving packets to/from the core network, either voice packet, or data packet.) which are the S1 and X2 handover procedures. The X2-handover procedure is normally used for the inter-eNodeB handover to balance network load and prevent interference. However, when there is no X2 interface between two eNodeBs, or if the source eNodeB has been configured to perform handover towards a particular target eNodeB via the S1 interface, then an S1-handover procedure will be triggered [12]. The S1-based handover procedure is used for communicating with non-3GPP specific access technologies such as CDMA2000/HRPD [13]. There are three phases involved in both S1 and X2 handover procedures which are preparation phase, execution phase, and completion phase [14, 15]. In the preparation phase, the UE needs to send measurement reports periodically to the source eNodeB [16]. Based on these reports, the source eNodeB will decide to which target eNodeB the UE should be handed over. Besides the measurement reports, other criteria are also considered by the source eNodeB before a control message is sent to the target eNodeB to prepare for the handover. Upon receiving the control message requesting to prepare for handover, the target eNodeB will prepare a buffer for the UE.

Once the preparation phase is completed, a handover command control message is sent by the source eNodeB to the UE in the execution phase to notify the UE that it is going to be handed over to another eNodeB. Upon receiving the message, the UE will disconnect itself from the source eNodeB and request for connection with the target eNodeB. At the same time, the source eNodeB forwards all packets of the UE to the target eNodeB. These packets are queued by the target eNodeB in the UE buffer. Once the UE has successfully connected to the target eNodeB, the target eNodeB transmits all the buffered packets of the UE followed by the incoming packets from the target gateway. The handover procedure moves to the completion phase after the UE sends to the target eNodeB a handover complete message that indicates this handover is completed.

The main purposes of the completion phase are to release all the resources used by the UE at the source eNodeB and to notify the upper layer to switch the path of the packet to the target eNodeB. Therefore, the target eNodeB needs to inform the source eNodeB to release all resources from the UE and the target MME to execute path switching to the target eNodeB, respectively.

## 3. HANDOVER ALGORITHMS

Three well known handover algorithms followed by the proposed algorithm in LTE system are discussed in this section.

### 3.1. LTE Hard Handover Algorithm

The LTE Hard Handover Algorithm, also known as "Power Budget Handover Algorithm", is a basic but effective handover algorithm consisting of two variables, handover margin (HOM) and Time to Trigger (TTT) timer [17]. A handover margin is a constant variable that represents the threshold of the difference in received signal strength between the serving and the target cells. HOM ensures the target cell is the most appropriate cell the mobile camps on during handover. A TTT value is the time interval that is required for satisfying HOM condition. Both HOM and TTT are used for reducing unnecessary handovers which is called "Ping-Pong effect". When a mobile is experiencing this effect, it is handed over from a serving cell to a target cell and handed back to original serving cell again in a small period of time [18]. This effect increases the required signaling resources, decreases system throughput, and increases data traffic delay caused by buffering the incoming traffic at the target cell when each handover occurs. Therefore effectively preventing unnecessary handovers is essential. TTT restricts the handover action from being triggered within certain time duration. A handover action can only be performed after the TTT condition has been satisfied. Figure 1 shows the basic concept of LTE hard handover algorithm. The received signal strength is called reference signal received power (RSRP) in dB (unless specified in dBm or dBW) in LTE system. When a mobile is moving away from the serving cell, the RSRP which the mobile receives from the serving cell will degrade as time increases. Meanwhile, the mobile will move towards the target cell, therefore the RSRP the mobile receives from the target cell will increase as time increases. A handover is triggered when the triggering condition (1) [19] and (2) are both satisfied, followed by the handover command.

$$RSRP_T > RSRP_S + HOM \qquad (1)$$

$$HOTrigger \geq TTT \qquad (2)$$

where $RSRP_T$ and $RSRP_S$ are the RSRP received from the target cell and the serving cell, respectively and HOTrigger is the handover trigger timer which starts counting when condition (1) gets satisfied.

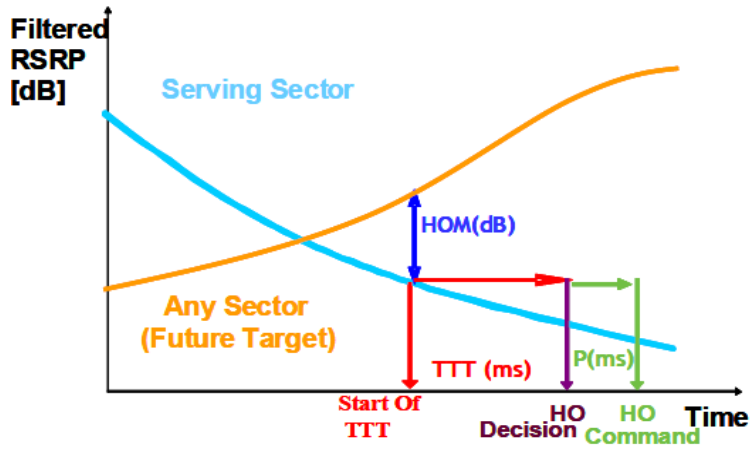

Figure 1.  LTE Hard Handover Algorithm [20]

## 3.2. Received Signal Strength based TTT Window Algorithm [21]

There are 3 steps involved in Received Signal Strength based TTT Window Algorithm. It collects required information during processing step, and then performs the comparison based on this information during decision step followed by the execution step.

$$RSS_F(nT_m) = \beta RSS(nT_m) + (1-\beta)RSS((n-1)T_m) \qquad (3)$$

$RSS_F$ is the filtered received signal strength ($RSS$, same as RSRP) measured at every handover measurement period ($T_m$) where $n$ and ($n$-$1$) is the $n^{th}$ and ($n$-$1$)$^{th}$ time instant, respectively. $\beta$ is a proposed fractional number called "forgetting factor" which can be expressed as follow:

$$\beta = \frac{T_u}{T_m} \qquad (4)$$

where $T_u$ is an integer multiple of $T_m$. A RSS comparison will be performed based on the following:

$$RSS_F(nT_u)_{TS} \geq RSS_F(nT_u)_{SS} + HOM \qquad (5)$$

*HOM* is a constant threshold value, $RSS_F(nT_u)_{TS}$ and $RSS_F(nT_u)_{SS}$ are the filtered RSS of the target sector (*TS*) and the filtered RSS of the serving sector (*SS*) at $(n\ T_u)^{th}$ interval, respectively.

This algorithm tracks the RSS value from each eNodeB and stores the instantaneous RSS value. Filtered RSS value at each instant is calculated using historical data (previously filtered RSS) by applying the forgetting factor variable. The closer the forgetting factor gets to 0, the higher the proportion that the current RSS depends on the filtered RSS in previous time instant. On the other hand, the closer the forgetting factor gets to 1, the higher the proportion that the current filtered RSS depends on the current RSS value. A handover decision will be made after (4) is satisfied for duration of whole $T_u$ window.

### 3.3. Integrator Handover Algorithm [17]

Integrator Handover Algorithm is a LTE handover algorithm proposed in 2008. The main concept is to make the handover decision by the historical signal strength differences. The idea of historical data is similar to what Received Signal Strength based TTT Window Algorithm has. There are 3 parts in integrator handover algorithm, RSRP difference calculation, filtered RSRP difference computation, and handover decision. The RSRP difference calculation is presented as following:

$$DIF_{s\_j}(t) = RSRP_T(t) - RSRP_S(t) \qquad (6)$$

where $RSRP_T(t)$ and $RSRP_S(t)$ represent the RSRP received from the target cell and serving cell at time $t$, respectively. $DIF_{s\_j}(t)$ is the RSRP difference of the user $j$ at serving cell $s$ at time $t$. The filtered RSRP difference computation can be written as following:

$$FDIF_{s\_j}(t) = (1-\alpha)FDIF_{s\_j}(t-1) + \alpha DIF_{s\_j}(t) \qquad (7)$$

where $\alpha$ is a proposed variable with constraint $0 \leq \alpha \leq 1$. $FDIF_{s\_j}(t)$ is the filtered RSRP difference value of user $j$ at serving cell $s$ at time $t$, and $DIF_{s\_j}(t)$ is the RSRP difference value calculated in (6). A filtered RSRP difference value will depend on the proportion between current RSRP difference and historical filtered RSRP difference in previous time instant by changing the $\alpha$ variable. The closer the $\alpha$ goes to 1, the higher the chance that filtered RSRP difference will have a heavier portion on the current RSRP difference calculated by (6). In the other way, the closer the $\alpha$ goes 0, the filtered RSRP difference will have a heavier portion on the previous historical filtered RSRP difference then on the current RSRP difference. Once the filtered difference has been computed, the handover decision will be made if the following condition is satisfied:

$$FDIF_{s\_j}(t) > FDIFThreshold \qquad (8)$$

*FDIFThreshold* is a constant value equivalent to HOM. If the filtered RSRP difference between any of target cell and serving cell is greater than this threshold, the handover decision will be triggered immediately. Please note ping-pong effect may occur due to lack of TTT mechanism involved in this algorithm.

### 3.4. LTE Hard Handover Algorithm with Average RSRP Constraint

LTE Hard Handover Algorithm with Average RSRP Constraint is proposed based on LTE Hard Handover Algorithm with an extra average RSRP condition for more efficient handover performance. The average RSRP can be calculated as following:

$$\text{RSRP}_{avgS\_j} = \frac{\sum_{n=1}^{N} \text{RSRP}_{S\_j}(nT_m)}{N} \qquad (9)$$

where $RSRP_{S\_j}(nT_m)$ is the RSRP received by user $j$ from serving cell $S$ at $n$-$th$ handover measurement period of $Tm$ and N is the total number of periods of duration $Tm$. An average RSRP of cell $S$ received by user $j$ ($RSRP_{avgS\_j}$) can be calculated by a sum of each $n$-$th$ handover measurement period ($Tm$) up to $N$ divided by $N$ times. An average RSRP constraint can be expressed as following:

$$\text{RSRP}_T(t) > \text{RSRP}_{avgS\_j} \qquad (10)$$

where $RSRP_T(t)$ is the current RSRP received from target cell $T$ and $RSRP_{avgS\_j}$ is the average RSRP computed from equation (9). The handover decision will be made by satisfying equation (10) followed by the same conditions as in LTE Hard Handover Algorithm listed below:

$$RSRP_T > RSRP_S + HOM \qquad (11)$$

$$HOTrigger \geq TTT \qquad (12)$$

A handover will be triggered if and only if equation (10), (11), and (12) are all satisfied. Please note $RSRP_{avgS\_j}$ will be reset to 0 each time due to serving cell changes when a handover is successfully performed.

The concept of this algorithm is to narrow down the possibility of handovers to minimize unnecessary handovers and ensure the channel quality of the target cell a user can have is not only higher than the current RSRP of serving cell with a certain threshold, but also better than the average RSRP received from the serving cell from the first handover measurement period till the last.

## 4. PERFORMANCE METRICS

The system performance of the four handover algorithms is evaluated on the basis of average handovers per UE per second, total system throughput, and total system delay. The average handovers per UE per second is the metric that is related to handover aspect whereas the system throughput and delay are network related performance metrics. Detailed descriptions of each metric are provided as below:

The average handovers per UE per second represents the number of handovers occurs during a simulation. It has the following expression:

$$\text{HO}_{avg} = \frac{\text{HO}_{Total}}{J \times T} \qquad (13)$$

where $HO_{avg}$ and $HO_{Total}$ are the average handovers per UE per second and total number of successful handovers, respectively and $J$ and $T$ are the total number of users and total simulation time, respectively. $HO_{Total}$ is incremented if and only if a handover is performed successfully. A successful handover is defined as a user has been handed over from source to target cell while maintaining the on-going data transmission. The cell throughput is defined as the total number of bits correctly received by all users per second. The cell throughput is measured at the eNodeB. It is mathematically expressed as:

$$\text{cell throughput} = \frac{1}{T} \sum_{j=1}^{J} \sum_{t=1}^{T} \text{tput}_j(t) \qquad (14)$$

where $tput_j(t)$ is the total size of correctly received packets (in bits) of user $j$ at time interval $t$, $T$ is the total simulation time and $J$ is the total number of users. The total system throughput is the sum of the 7 cells throughput expressed below:

$$\text{throughput}_{\text{Total}} = \sum_{c=1}^{C} \text{cell throughput}_c \qquad (15)$$

where *cell throughput$_c$* is the individual cell throughput of cell $c$ calculated from equation (14) and $C$ is the total cells in the simulation. System delay is defined as average system Head-of-Line (HOL) delay or queuing delay. A HOL delay is defined as the time duration from the HOL packet's arrival time at the eNodeB buffer to current time. It can be expressed in the following equation:

$$\text{Cell Delay} = \frac{1}{T}\sum_{t=1}^{T}\frac{1}{J}\sum_{j=1}^{J} W_j(t) \qquad (16)$$

where $J$ is the total number of users within the cell, $T$ represents the total simulation time, and $W_j(t)$ denotes the HOL delay of user $j$ at time $t$. The total system delay is the sum of the 7 cells delay expressed below:

$$\text{Packet Delay}_{\text{Total}} = \sum_{c=1}^{C} \text{Packet Delay}_c \qquad (17)$$

where *Packet Delay$_c$* is the individual cell throughput of cell $c$ calculated from equation (16) and $C$ is the total cells in the simulation.

## 5. SIMULATION ENVIRONMENT AND DESCRIPTIONS

The performance of four handover algorithms previously discussed are evaluated, optimized and compared using a C++ platform computer simulation which simulates the downlink LTE system consisting of a 7-hexagonal-cell scenario of 5 MHz bandwidth with 25 RBs and 2 GHz carrier frequency based on [22, 23] containing 100 users. Users are uniformly distributed within the rectangle area as shown in Figure 2.

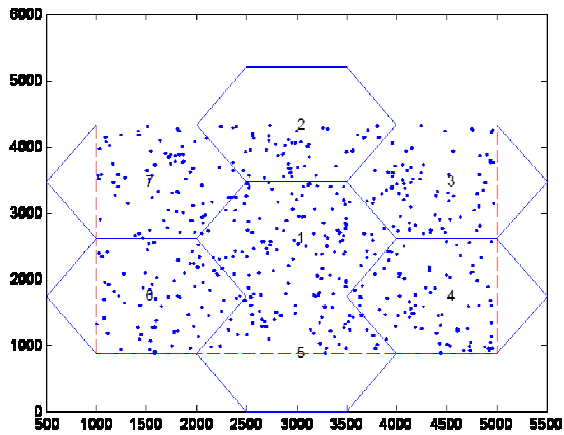

Figure 2. Simulation Environment

Each eNodeB is located at the centre of each cell with 100m radius and it is assumed that equal transmit power (43.01 dBm total eNodeB transmit power) is used on each RB. Each UE is constantly moving at a fixed speed and the speed is varied under three different scenarios (i.e. 3 km/hr, 30 km/hr or 120 km/hr scenarios). Direction of each UE is randomly chosen between 0 to $2\pi$, initially and stays constant in throughout its session.

Users are wrapped around whenever they reach the red rectangle edge. [24]. The Cost-231 HATA model for an urban environment [25, 26] is used to compute pathloss. A Gaussian log-normal distribution with 0 mean and 8 dB standard deviation [27] is used for modelling shadow fading. A Non-frequency selective Rayleigh fading [28] is used to model the radio propagation channel. A user that has data to receive estimates its instantaneous Signal-to-Interference-Noise-Ratio (SINR) on each RB, converts it into a CQI value with the target block error rate (BLER) to be less than 10 % [7] and reports each CQI value to the source eNodeB. It is assumed in this paper that the CQI reporting is performed in each TTI of 1 millisecond and on each RB. A total of 16 Channel Quality Information (CQI) levels as defined in [7] are used. The equations for generating BLER curves in [29] are used for modeling the performance of turbo codes in Rayleigh fading channel. The Hybrid Automatic Repeat Request (HARQ) technique in [30] is adopted to recover wireless transmission errors. The CQI and HARQ reporting are modeled error free with 3 ms CQI delay and 4 ms HARQ (ACK/NACK) delay. The maximum number of retransmissions is limited to 3. Round-Robin packet scheduler is chosen for a fair transmission opportunity for all users and a 50 milliseconds interval is set for each user's measurement report for handover decision. A shorter simulation time of 1000 and longer simulation time of 10000 milliseconds are used for performance optimization and handover algorithms performance comparison, respectively. System parameters used in the simulation are given in Table 1.

Table 1.  Simulation Parameters

| Parameters | Values |
|---|---|
| Cellular layout | Hexagonal grid, wrap around (reflect), 7 cells |
| Carrier Frequency | 2 GHz |
| Bandwidth | 5 MHz |
| Number of RBs | 25 |
| Number of sub-carriers per RB | 12 |
| Sub-carrier Spacing | 15 kHz |
| Path Loss | Cost 231 Hata model |
| Shadow fading | Gaussian log-normal distribution |
| Multi-path | Non-frequency selective Rayleigh fading |
| Packet Scheduler | Round Robin |
| Scheduling Time (TTI) | 1 ms |
| Data Traffic | 1 Mbps Constant Rate |
| User | 100 |
| User's position | Uniform distributed |
| User's direction | Randomly choose from [0,2$\pi$], constantly at all time |
| Simulation time | 1000 ms for optimization 10000 ms for performance evaluation |
| RSRP sampling timer interval | 50 ms |

The optimization parameters are determined by comparing the new so-called *OptimizeRatio* value which is a ratio calculated by total system throughput over the average number of handovers. *OptimizeRatio* can be computed as following:

$$\text{OptimizeRatio}_{(HOA,\text{Speed})} = \frac{\text{ST}_{(HOM,TTT)}}{\text{ANOH}_{(HOM,TTT)}} \qquad (18)$$

where *HOA* indicates the handover algorithm, *Speed* is the corresponding speed in each scenario. *ST* and *ANOH* are the total system throughput of sum of 7 cells and the average number of handover per UE per second, respectively. *TTT* will be replaced by $\alpha$ or $\beta$ factor when Integrator Handover Algorithm or Received Signal Strength based TTT Window Algorithm is selected.

Table 2.  Optimization Parameters

| Parameters | Values |
|---|---|
| Handover Algorithm (HOA) | 1: LTE Hard Handover Algorithm<br>2: Received Signal Strength based TTT Window Algorithm<br>3: Integrator Handover Algorithm<br>4: LTE Hard Handover Algorithm with Average RSRP Constraint |
| TTT | {0,1,2,3,4,5} millisecond |
| HOM | {0,1,2,3,4,5,6,7,8,9,10} dB |
| UE Speed | {3,30,120} km/hr |
| $\alpha / \beta$ | {0.25, 0.5, 0.75, 1} |

Table 2 outlined the LTE Standard Hard Handover Algorithm, Received Signal Strength based TTT Window Algorithm, Integrator Handover Algorithm and proposed LTE Hard Handover Algorithm with Average RSRP Constraint are referred as HOA 1, HOA 2, HOA 3, and HOA 4 respectively, in the following discussions. The maximum TTT value of the RSRP sampling timer interval is assumed to be 10% in the simulation. The range of the HOM (*FDIFThreshold*) value and α (β) factor are similar to that given in [17]. The highest *OptimizeRatio* value leads to a set of optimized parameters of the selected handover algorithm under a specific speed condition by having maximizing the total system throughput and minimizing the unnecessary average number of handovers per UE per second. Note that, an *ANOH* value equals to 0 is replaced to 0.5 to avoid numerical calculation error.

# 6. SIMULATION RESULTS

Optimization of four handover algorithms is discussed in this section followed by the performance evaluation and comparison under 3 speed scenarios.

## 6.1. Optimization

The *OptimizeRatio* results in Figure 3 are calculated using equation (18) by having input sets as HOA 1 and UE speeds equal to 3, 30, and 120 km/hr with changing HOM value from 0 to 10 and TTT value from 0 to 5. The highest bar in each speed scenario in Figure 3 indicates the highest *OptimizeRatio* value in each simulation and it refers to HOM and TTT equal to 10 and 5 in 3 km/hr scenario, HOM and TTT equal to 6 and 5 in 30 km/hr scenario, and HOM and TTT equal to 7 and 5 in 120 km/hr scenario, respectively.

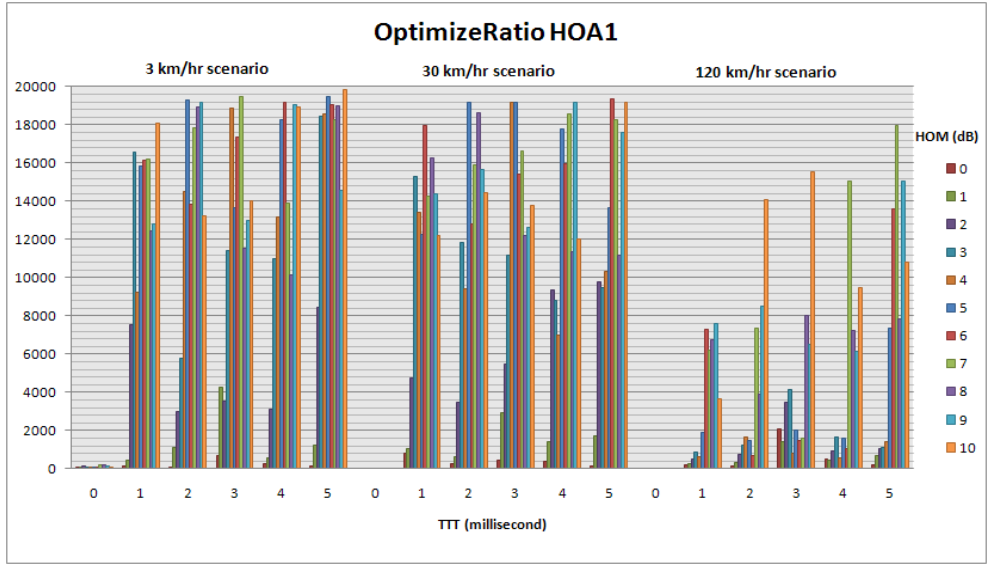

Figure 3.  *OptimizeRatio* in HOA 1

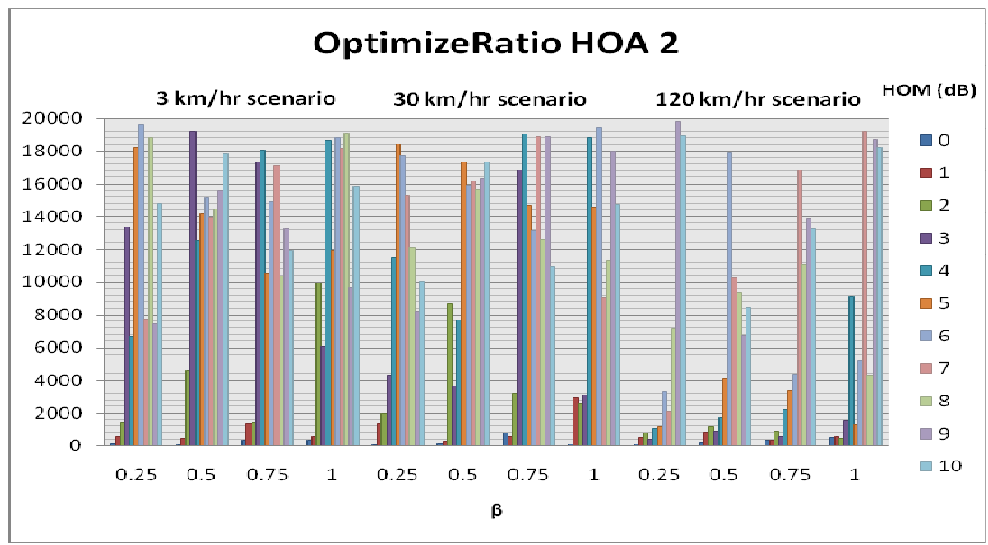

Figure 4.  *OptimizeRatio* in HOA 2

Figure 4 demonstrates the *OptimizeRatio* in HOA 2 with 3 speed scenarios. The highest *OptimizeRatio* value in 3 km/hr scenario, 30 km/hr scenario, and 120 km/hr scenario, are β and HOM equal 0.25 and 6, β and HOM equal 1 and 6, and β and HOM equal to 0.25 and 9, respectively.

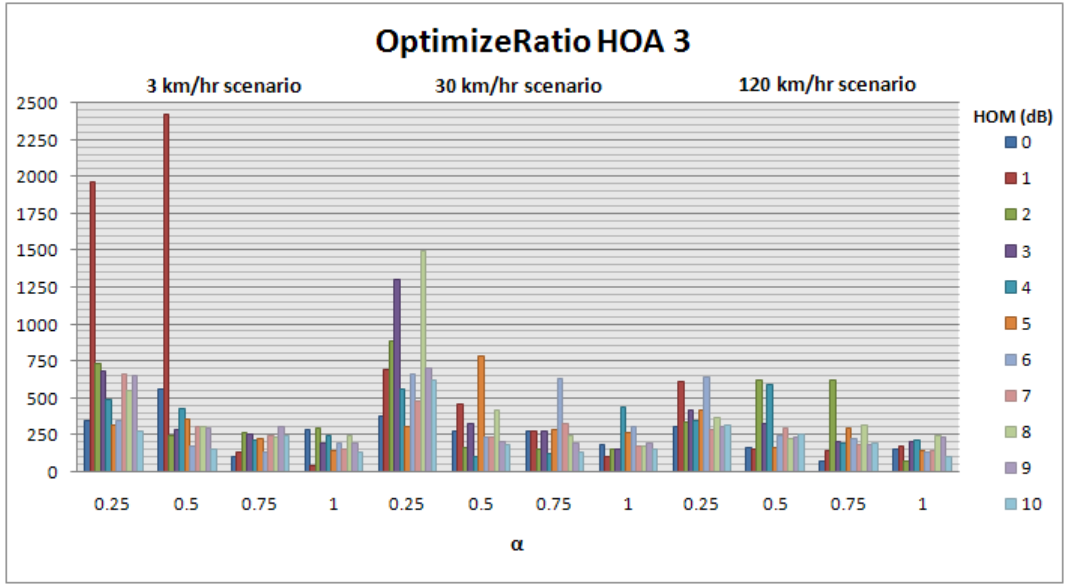

Figure 5. *OptimizeRatio* in HOA 3

The highest *OptimizeRatio* value in HOA 3 can be seen in Figure 5 as α and HOM equal 0.5 and 1, 0.25 and 8, and 0.75 and 2 in 3 km/hr, 30 km/hr, and 120 km/hr speed scenario, respectively.

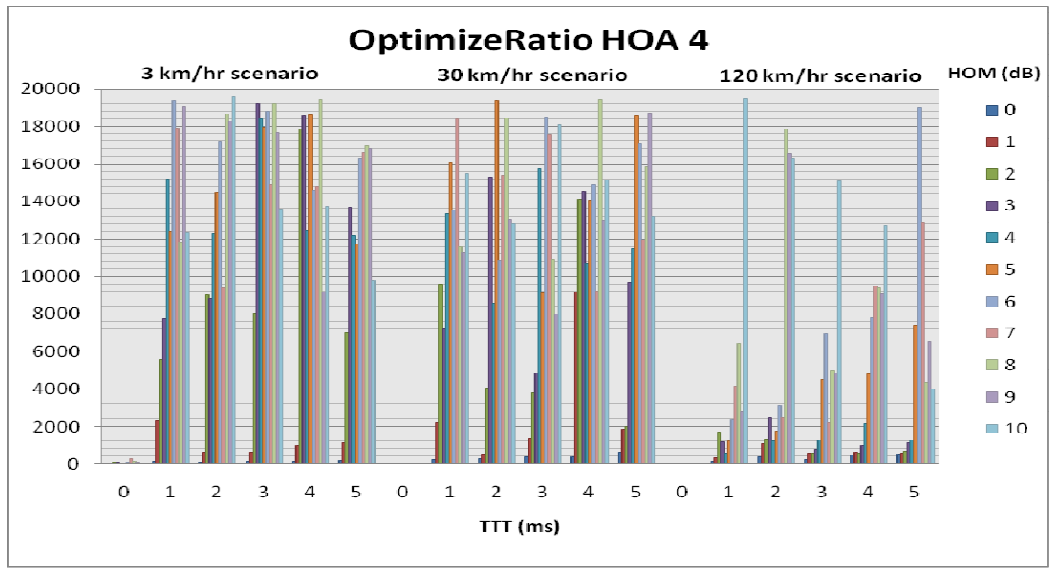

Figure 6. *OptimizeRatio* in HOA 4

A set of optimized parameters of HOA 4 is determined in Figure 6 as HOM and TTT equal 10 and 2, 8 and 4, and 10 and 1 in scenario under speed of 3, 30, 120 km/hr scenario, respectively.

Table 3 shows a summarized result of the optimized parameters for each handover algorithm for varying speed scenarios.

Table 3.  Optimized Parameters

| Speed [km/hr] | HOA 1 | HOA 2 | HOA 3 | HOA 4 |
|---|---|---|---|---|
| 3 | [HOM, TTT] = [10, 5] | [HOM, β] = [6, 0.25] | [HOM, α] = [1, 0.5] | [HOM, TTT] = [10, 2] |
| 30 | [HOM, TTT] = [6, 5] | [HOM, β] = [6, 1] | [HOM, α] = [8, 0.25] | [HOM, TTT] = [8, 4] |
| 120 | [HOM, TTT] = [7, 5] | [HOM, β] = [9, 0.25] | [HOM, α] = [6, 0.25] | [HOM, TTT] = [10, 1] |

The remaining results of the performance comparisons are based on the optimized parameters as listed in Table 3.

## 6.2. Performance Evaluation and Comparison

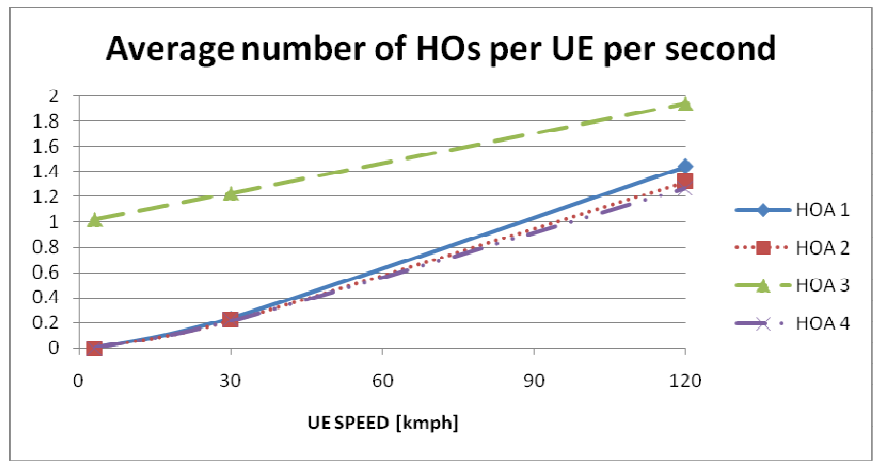

Figure 7.  Average Number of Handovers per UE per Second of 4 handover algorithms

Figure 7 shows the average number of HOs per UE per second under the four handover algorithms with increasing UE speeds. Since the HOA 3 does not implement TTT mechanism, it can be seen in the figure that, with increasing UE speeds, the average number of handovers per UE per second under the HOA 3 is significantly higher as compared with all the other three handover algorithms. All three handover algorithms (i.e. HOA 1, HOA 2, and HOA 4) achieve a similar average number of handovers per UE per second at a lower user speed whereas almost comparable average number of handovers per UE per second achieve under the HOA 2 and HOA 4 respectively at a higher UE speed. The result shows that the HOA 4 has a sum of average number of handovers per UE per second of 3 speed scenarios as 1.49 which is less than 1.68, 1.54, and 4.19 of HOA 1, HOA 2, and HOA 3, respectively. Furthermore, the sum of average number of handovers per UE per second in proposed HOA 4 is effectively reduced up to 35.56% when compared with the HOA 3.

Figure 8 shows the total system throughput of 7 cells under 4 handover algorithms with increasing UE speeds. A higher total system throughput value implies a higher system performance a handover algorithm offers. The figure demonstrates that HOA 3 has a highest total system throughput as 77.2496 Mbps at 3 km/hr due to users frequently handover for cells which have better channel quality at low speed but the total system throughput drops gradually to 55.9141 and 41.976 Mbps at speed 30 and 120 km/hr respectively, due to increase in number of handovers resulting in network congestion and therefore the drop in system performance.

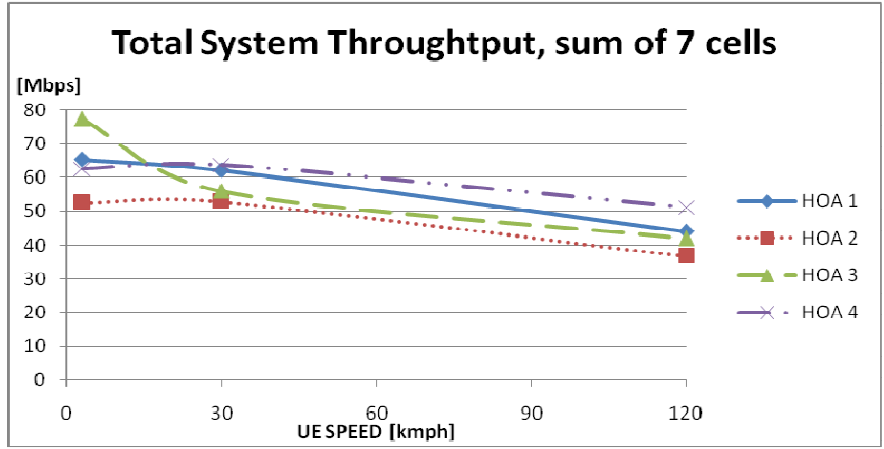

Figure 8.  Total System Throughput, sum of 7 cells of 4 handover algorithms

A long TTT window ($T_u$ = 100 ms) delays the time to execute the handover therefore resulting with the HOA 2 having a lowest total system throughput in all speed scenarios. A sum of total system throughput of HOA 4 in all speed scenarios of 177.4205 Mbps is the highest value compared with 171.3447, 141.8809, and 175.1397 Mbps of  HOA1, 2, and 3, respectively, furthermore, the sum of total system throughput of HOA 4 has a 3.55% , 25%, and 1.302% performance improvement of HOA 1, HOA 2, and HOA 3, respectively.

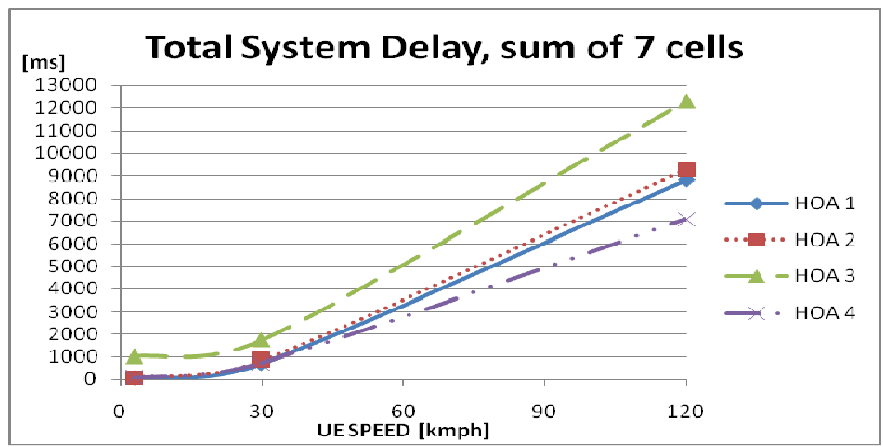

Figure 9.  Total System Delay, sum of 7 cells of 4 handover algorithms

Figure 9 demonstrates the total system delay of 4 handover algorithms in 3 speed scenarios. Since the handover is more likely to occur frequently as the speed increases, this results with increasing system delay under all handover algorithms being evaluated. The HOA 3 has a slightly higher delay due to lack of TTT mechanism at all speed scenarios as compared with the other three handover algorithms. The HOA 4 has the smallest total system delay at all speed scenarios (i.e. 63.1917, 742.917, and 7082.12 ms at 3, 30, 120 km/hr, respectively). The sum of total system delay of HOA 1, HOA 2, HOA 3, and HOA 4 in all speed scenarios are 9611.00, 10214.45, 15048.69 and 7888.23 ms, respectively. This result shows that the sum of total system delay of HOA 4 outperforms the sum of total system delay of HOA 1, 2, and 3 in all speed scenarios by 17.93%, 22.77%, and 47.58% less delay, respectively.

## 7. CONCLUSIONS

A new handover algorithm is proposed in this paper and its impact for a number of optimized handover parameters under the downlink LTE system is evaluated. The performance of the proposed algorithm is compared with three well known handover algorithms under different UE speed scenarios. It shown via computer simulation that the proposed handover algorithm can effectively reduce the average number of handovers per UE per second up to 35.56% when compared with Integrator Handover Algorithm. Moreover, the total system throughputs under the proposed handover algorithm are 3.55%, 25%, and 1.302% higher as compared to the LTE Hard Handover, RSS Based TTT Window and Integrator Handover Algorithms, respectively. Similarly, the proposed handover algorithm is able to maintain a lower system delay when compared with the other three well known handover algorithms (i.e. 17.93%, 22.77%, and 47.58% reductions when compared with LTE Hard Handover, RSS based TTT Window and Integrator Handover Algorithms, respectively). Future studies include evaluating the performance of the proposed handover algorithm under different wireless scenarios taking QoS requirements of multimedia services under consideration.

## Authors

Cheng-Chung Lin is currently a PhD Candidate in the Faculty of Engineering and Information Technology, University of Technology, Sydney (UTS), Australia. He received a graduate certificate (GradCert) in advanced computing (2006) and a Master of Information Technology in internetworking in School of Computer Science and Engineering from University of New South Wales, Australia (2007). His current research interests focus on handover and packet scheduling in radio resource management for the future wireless networks.

Kumbesan Sandrasegaran (Sandy) holds a PhD in Electrical Engineering from McGill University (Canada) (1994), a Masters of Science Degree in Telecommunication Engineering from Essex University (UK) (1988) and a Bachelor of Science (Honours) Degree in Electrical Engineering (First Class) (UZ) (1985). He was a recipient of the Canadian Commonwealth Fellowship (1990-1994) and British Council Scholarship (1987-1988). He is a Professional Engineer (Pr.Eng) and has more than 20 years experience working either as a practitioner, researcher, consultant and educator in telecommunication networks. During this time, he has focused on the planning, modeling, simulation, optimisation, security, and management of telecommunication networks.

Huda Adibah Mohd Ramli is currently a PhD candidate in the Faculty of Engineering and Information Technology, University of Technology, Sydney (UTS), Australia. She received a M.Sc. in Software Engineering from University of Technology Malaysia (2006) and B.Eng. in Electrical and Computer Engineering from International Islamic University Malaysia (2003). Her current research interests focus on radio resource management for the future wireless IP networks.

Riyaj Basukala received his Masters of Engineering (Telecommunication Engineering) Degree from UTS in 2009 and Bachelors of Engineering in Electronics and Communication from Tribhuwan University, Nepal in 2005. Currently he is involved in working as a Research Assistant in Centre for Real Time Information Networks (CRIN) in UTS focussing mainly on developing simulation tools for simulating radio resource management in various radio interface technologies and also few other communication-based research projects.